%% file: main-paper.tex
\documentclass{article}

\usepackage[utf8]{inputenc}
\usepackage{geometry}
\usepackage{graphicx}
\usepackage{algorithm}
\usepackage{algorithmicx}
\usepackage{algpseudocode}
\usepackage{caption}
\usepackage{subfig}
\usepackage{amsmath}
\usepackage{multirow}
\usepackage[acronym]{glossaries}

\algdef{SE}[DOWHILE]{Do}{doWhile}{\algorithmicdo}[1]{\algorithmicwhile\ #1}

\title{An Interleaving Hybrid Consensus Protocol}
\author{
  Yao Sun\\
  \texttt{yao@theoan.com}
  \and
  Aayush Rajasekaran\\
  \texttt{aayush@theoan.com}
}

\date{November 20, 2019}

\begin{document}

\maketitle

\begin{abstract}
    We introduce Unity Interleave, a new consensus algorithm for public blockchain settings.
    It is an eventual consistency protocol merging the Proof-of-Work (PoW) and Proof-of-Stake (PoS)
    into a coherent stochastic process. It builds upon research previously done for the Unity
    protocol, improving security while maintaining fairness and scalability.
\end{abstract}

\section*{Acknowledgements}

This specification was produced in collaboration with the entire R\&D staff at the Open Foundation, with major contributions from Ali Sharif, Ge Zhong, Jeff Disher, Shidokht Hejazi, Sam Pajot-Phipps, Yunfei Zha and Alexandra Roatis.

\newpage
\tableofcontents
\newpage


\input{sections/1introduction-related.tex}
\input{sections/2background.tex}
\input{sections/3definition.tex}
\input{sections/4analysis.tex}

\input{sections/5results.tex}
\input{sections/6future-works-conclusion.tex}


\pagebreak
\bibliographystyle{alpha}
\bibliography{paper}

\end{document}

%% file: sections/1introduction-related.tex
\section{Introduction}
\label{intro}

The idea of \emph{Proof of Work (PoW)} originated in the work of Dwork and Naor
in the 1990s\cite{dwork-naor}. Used in the majority of blockchain protocols today, including
Bitcoin, the idea is that a block can only be considered valid if it includes
a solution to some cryptographic puzzle that is hard to solve, but easy to verify. 
Block producers on a PoW network are often called \emph{miners}, and 
the security of a PoW network relies on the difficulty of finding solutions
to these puzzles. For more on PoW consensus, see \cite{nakamoto2008}.

A PoW network is considered to be secure if a large percentage of the hash power
solving these puzzles is controlled by honest actors (usually stated as 51\% of the hash power,
though the number can be as high as 66\% \cite{selfish_mining}). 
The rise of large mining pools \cite{aliaga2018}, as well as the new possibility of renting
hash power makes it easier for a single entity to acquire large quantities of hash power,
especially on less well-secured networks. This revokes previous assumptions around
the security of blockchains, leaving smaller non-ASIC networks especially vulnerable \cite{sinnige2018}.

An alternative approach known as \emph{Proof-of-Stake (PoS)} encompasses a broad spectrum of
consensus protocols that share the common attribute in utilizing the \textit{stake} a user has in a network \cite{BentovGM14}.
Block producers on a PoS network are often called \emph{stakers}, and 
this stake is usually some economic asset that lives on the network itself.
The security of networks built on PoS consensus relies on incentivising block producers
to support the network they are on (since they have assets on the network), and by 
punishing behaviour that can be shown to be dishonest (by confiscating the assets of
a dishonest actor). PoS networks are vulnerable to a different set of attacks than PoW networks,
see \cite{brown2018formal} for more.

In this paper, we present a new consensus protocol that combines the ideas
of PoW and PoS. This protocol, which we call ``Unity-Interleave'', is a modification
of the protocol described in \cite{wu2019unifying}.
The approach is to simply alternate between PoW and PoS blocks; a miner
cannot produce a PoW block on top of another PoW block, and the same is
true of stakers and PoS blocks. We strive to achieve the benefits of both PoW and
PoS protocols, while ameliorating their vulnerabilities.

\textbf{Paper Layout.} In \S\ref{background} we go over some of the background of blockchain technology.
\S\ref{unity} defines the protocol itself in detail, and \S\ref{analysis} 
analyses the protocol's resistance to various attacks. Finally, in \S\ref{results},
we present the results of some simulations that prove our protocol works as desired.

%% file: sections/2background.tex
\section{Background}
\label{background}

Assumed a simplified model for a blockchain consisting of \textit{blocks} and \textit{transactions}. Assume each transaction is some arbitrary unique token (e.g. a unique 32-byte hash), and each block is a data structure that
bundles an ordered series of transactions. As with earlier blockchain protocols, blocks in this scheme must reference a valid, previously published unique block (commonly referred to as a parent). See \cite{wood2014ethereum} for a more complete model for a blockchain. 

The purpose of Proof of Work (\cite{nakamoto2008}) is to provide an easily verifiable cryptographic proof that a certain amount of work has been done. It is used by the protocol to generate the notion of \textit{difficulty}, in that proofs generated from a block with higher difficulty require more computational resources than that of a lower difficulty. Therefore, difficulty is a measure of total \textit{work} done to produce a block. This numerical value in turn can be utilized to determine the link of blocks that represent the chain with the \textit{most work} via summation, among other approaches (this paper included) in agreement on a canonical chain. This in turn enforces agreement on a canonical ordering of transactions, which forms the basis for all blockchain interactions.

%% file: sections/3definition.tex
\section{Protocol Description}
\label{unity}

We keep and maintain the notion of difficulty provided by Proof of Work, as well as the general goal of the consensus protocol, which is agreement on canonical ordering of blocks. We introduce an alternative block generation scheme utilizing \textbf{stake} or \textbf{voting power} while maintaining the constructs own notion of difficulty. To facilitate this, we introduce a mechanism to obtain \textbf{stake} in the network by locking up network currency associated with each account. To maintain a split between the two types of blocks, we introduce a new rule that enforces the \textit{interleaving} behaviour, in that blocks must reference a parent of the opposing type. Finally, we introduce a fork choice rule that determines agreement on a canonical chain based upon an intermediate unit derived from the difficulty of both block generation algorithms. We show that this approach leads to an increase in security against certain attacks \S\ref{analysis}.

In order to forge a PoS block, a staker must have already registered some stake in a previous block. The exact mechanism of how stake is registered is an implementation detail; for a summary of how the Open Application Network manages staking, see \cite{unity_design_spec}, and \S\ref{staking_mechanism} for a very brief overview. A PoS block has some waiting time associated with it, and will only be accepted by the network once this wait time has elapsed (measured from the parent PoW block). Every staker has a random wait time calculated for each block they try to produce; however, the expected wait time is inversely proportional to the amount a staker has staked. In short, the more one stakes, the less their wait times are expected to be. The percentage of blocks a staker should expect to win is their percentage of the total stake in the network.

\subsection{Staking}
\label{staking_mechanism}

Staking in this context refers to \textbf{locking} ones assets in return for voting power (alternatively known as stake). In our model, the act of locking ones assets can be done immediately (for example in the next block), the act of \textbf{unlocking} ones assets however is associated with a waiting period before one can re-acquire the assets (for example, the assets can be re-acquired after $x$ blocks). We ignore concepts of transferring and delegating (in the context of the protocol) as described in \cite{unity_design_spec}.

\subsection{PoW Block Generation}
\label{pow_block_generation}

Miners performing Proof of Work to generate the next block use a scheme similar to the introduced by Nakamoto in \cite{nakamoto2008}. Assuming a mining difficulty parameter ($d_w$), miners solve the following cryptographic puzzle,

\begin{equation}
\label{eq:pow_puzzle_equation}
H(b) \leq 2^{256} / d_w.
\end{equation}

The hash function ($H()$) is a deterministic random oracle that returns a random number in $N_{256}$, given an arbitrary byte array $b$ as input ($b$ is usually referred to as the \textit{nonce}). As the cryptographic puzzle is independent of any miner, the probability of a miner proposing a block is proportional to their computational power (number of permutations of $H()$ the miner can evaluate per unit time). The difficulty parameter ($d_w$) is used by the network to stabilize it's block production rate in the presence of dynamic hash power (see \S\ref{difficulty_adjustment}).

\subsection{PoS Block Generation}
\label{pos_block_generation}

The generation of PoS blocks is largely based on the Nxt forging algorithm \cite{popov2016probabilistic}. 

A PoS block has a field called \emph{seed}. A PoS block's seed is based on the seed of the previous PoS block in the chain, which is always the block's grandparent. The seed of block $n + 2$ is the block producer's signature upon the seed of block $n$ (i.e. in order to produce block $n+2$, the staker must obtain the seed for this block by signing the seed of the last PoS block (block $n$)).  

Any time a staker is trying to produce a PoS block, there is a certain waiting time that must elapse before the block is valid. This wait time is measured from the timestamp of the parent block. The wait time is determined by three factors: the new block's seed, $s$, the staker's total stake, $V$, and the difficulty of the block, $d_s$. The difficulty is calculated as described in \S\ref{difficulty_adjustment}. The wait time, $\Delta$, is inversely proportional to $V$, and directly proportional to $d_s$. The seed, $s$, provides some randomness to this calculation. We calculate the wait time as,

$$\Delta = \frac{d_s \cdot \log (\frac{2^{256}}{H(s)})}{V},$$

where $H$ is some hash function producing a 256-bit output.

The timestamp of a PoS block must be greater than or equal to the parent block's timestamp plus this $\Delta$. Honest nodes should thus only accept PoS blocks once the waiting time has elapsed .Further more, honest nodes must reject blocks with timestamps that are greater than their system time, so as to not accept block \emph{in the future}.

\subsection{Fork choice rule}
\label{fork}

The \emph{fork choice rule} is the aspect of a consensus protocol that decides between two or more chains of valid blocks to decide what the ``canonical'' chain or ``main'' chain is. The fork choice rule for the protocol is to always choose the chain that maximises total difficulty, where total difficulty is simply the sum of the difficulties of all the blocks in the chain. Therefore, given a set of chains $D$, we denote the $i$th chain as $d_i$, and we denote $td_{s, d_i}$ and $td_{w, d_i}$ to be the total staking and work difficulties (respectively) for the $i$th chain. We recognize the main chain as the one that has in its prefix the genesis block, and that satisfies,

\begin{equation}
    M = \{td_{s, d_i} + td_{w, d_i} > td_{s, d_j} + td_{w, d_j} \forall j \neq i | d_i \in D \}
\end{equation}

In \S\ref{double_spend_attack} we examine why this fork choice rule is beneficial, compared to the one depicted in \cite{wu2019unifying}.

\subsection{Difficulty Adjustment}
\label{difficulty_adjustment}

We begin with the presumption that block arrival times for fast acting difficulty adjustment algorithms can be modelled using an exponential distribution. Note that this is not true in the case of a blockchain with a lagging difficulty adjustment algorithm such as Bitcoin \cite{block_arrivals}. This difficulty algorithm is applied identically to all blocks in the chain (i.e. both PoW and PoS blocks). Given that $\Delta$ represents the block arrival rates, the difficulty of the ($n+1$)th block (denoted $d_{n+1}$) is related the to the difficulty of the $n$th block (denoted $d_n$) via:

\begin{equation}
    d_{n+1} = \begin{cases}
        \frac{d_n}{1+\alpha}, & \Delta > -\frac{\log(1/2)}{\lambda} \\
        d_n, & \Delta = -\frac{\log(1/2)}{\lambda} \\
        d_{n} \cdot (1 + \alpha), & \Delta < -\frac{\log(1/2)}{\lambda} \\
    \end{cases}
\end{equation}

Where $\lambda$ represents the rate parameter for an exponential distribution. The $\alpha$ controls the learning rate, which further determines the responsiveness of the algorithm. The boundary value of $-\frac{\ln (0.5)}{\lambda}$ is calculated by solving,

\begin{equation}
    \label{eq:boundary}
    1-e^{(-\lambda x)}=0.5,
\end{equation}

where the left-hand side is the cumulative distribution function (cdf) of an exponential random variable. Where equation \ref{eq:boundary} refers to the time at which the probability of a block arriving or having already arrived is half. To give an example, in the live network we design our expected block time to be centered at 10 seconds, therefore we design the boundary to target an exponential distribution with a mean of 10. This gives us $\lambda^{-1}=10$. Solving $x$ for such a lambda gives us a boundary time of $x \approx 6.9314$.

Intuitively, this creates a negative feedback loop that stabilizes the system around 10 seconds. If the average difficulty is too high, it is reflected in the system by an increase in block time (on average), in which case the protocol responds by decreasing the difficulty. In the same vein, difficulty is adjusted upwards when the average block time is too low.

%% file: sections/4analysis.tex
\section{Analysis}
\label{analysis}

In this section, we assess the protocol's resistance to various attacks. We claim that in most cases, the protocol behaves no worse than a pure PoW network. 

\subsection{Double-spend attack}
\label{double_spend_attack}

The \textit{double-spend attack} is the well-known, but somewhat poorly named, ``51\% attack'' \cite{selfish_mining}. In the simplest version of this attack, a malicious agent sends some transaction to the network to reap an off-chain benefit (say she sells a large amount of her on-chain assets on an exchange), and then releases a side-chain in which that transaction never occurred. If this side-chain becomes the main-chain, she has 
exploited the network.

Large parts of the protocol were designed with the specific goal of frustrating this attack. The interleaving of PoW and PoS blocks serves to make this attack much harder to launch.

We model this attack by letting the average difficulty of PoW blocks on the chain be $d_w$, and the average difficulty of PoS blocks $d_s$. We let the attacker have $k$ times the hashpower and $l$ times the stake of the honest block producers on the network. Once the difficulty has adjusted on the attacker's side-chain, she will be producing PoW blocks of average difficulty $kd_w$, and PoS blocks of average difficulty $ld_s$. She should expect to win if her chain grows heavier 
than the honest-chain in the long-run. Mathematically, this means she needs $$kd_w + ld_s > d_w + d_s$$

We can easily observe that if $k < 1$ and $l < 1$, the attacker is not expected to win, and if \textit {both} $k>1$ and $l>1$, the attacker is always expected to win. Therefore, 51\% of at least one
domain is necessary to expect to win, and 51\% of both is sufficient to expect to win. 

The intermediate case is when the attacker has a majority of one domain, but not the other. We let the attacker have $k>1$ and $l<1$ (the other case is symmetric). We can rewrite the winning condition as 
$$k > 1 + (1-l)\frac{d_s}{d_w}$$

Note that the difficulty values $d_w$ and $d_s$ are just indications of how hard it was to produce a block, and therefore can be controlled to some extent by the designers of a protocol. This gives us
the nice property that by manipulating these difficulty values, we can make the attack harder to launch in one direction than the other. On the OAN, for instance, we believe that it will be easier for attackers to acquire a majority of the hash power ($k > 1$) than a majority of the stake. Accordingly, we set the typical
$d_s$ to be several orders of magnitude larger than $d_w$ (roughly $10^{14}$ times larger). The result is that an attacker with, say, 9/10ths the stake of the honest network still needs to have over $10^{10}$ times as much hash power as the honest work. The converse effect is that an attacker with a majority of the stake needs very little hash power to dominate the network, but we believe acquiring a majority of the stake to be infeasibly difficult.

\subsection{Nothing-at-Stake}

The  \textit{Nothing-at-Stake} problem arises in pure PoS blockchains when stakers try to produce blocks on every branch they see \cite{brown2018formal}. Since there is no cost to PoS block production, this is a profit-maximising strategy for PoS block forgers. This can lead to a fragile network with many branches and side-chains. 

Our protocol is not susceptible to this problem since the miners will resolve the branch; of several valid PoS blocks at the same level, whichever one has a PoW block built on top of it first will likely be included into the canonical chain.

\subsection{Stake Grinding}

A PoS block producer is said to be \textit{stake grinding} if, in the production of a block, they can somehow increase the chances that they produce the next block too \cite{buterin_randomness}. In our protocol, the randomness that determines block production is entirely determined by the seed of the previous PoS block, and the private key of the block producer \S\ref{pos_block_generation}.

At every block an attacker will try to brute force compute an account that generates a lower delta for the next $k$ blocks, then attempt to transfer ownership of coins over to that account. The attack would proceed as follows: assume the attacker knows that they can produce block $n$. In the time that the attacker has before he has to announce his block ($\Delta$), the attacker can select an account (via time-bounded brute-force search) which minimizes the $\Delta$ for the next PoS block. The attacker then includes a transaction into block $n$ which transfers the attacker's stake to the selected account. When it's time to produce the next block, assuming the attacker has the smallest $\Delta$ and is the block producer at this round as well, the attacker repeats this behavior as long as they can sustain this \emph{streak} of discovering favorable accounts which yield small $\Delta$s. 

Notice that this attack is made feasible by the instantaneous transfer of stake from one account to another. Therefore, if we enforce a lockout period (of $x$ blocks) before a staker is allowed to transfer stake to another account, the attacker must now win at-least $x+1$ blocks consecutively under the same account in order to successfully launch this attack. Therefore, the probability of success of the stake grinding attack can be reduced to the probability for an honest staker of winning $x+1$ blocks consecutively.

Let $s_i$ denote the absolute stake controlled by the $i$th staker and let $p_i$ denote the probability that the next block is proposed by the $i$th staker. It is easy to see that for PoS blocks, the probability of winning the next block for an honest staker is denoted by: 

$$p_i=s_{i}/\sum_i{s_i}$$

Assuming that $\sum_i{s_i}$ is constant for the next $x$ blocks, it follows that the $i$th staker's probability of winning $x+1$ blocks consecutively is denoted by

\begin{equation}
    \label{eq:attacker_success}
    P(\texttt{wining x consecutive blocks})=p_i^{x}
\end{equation}

This equation implies that the probability that an honest staker can produce $x$ consecutive blocks diminishes by a factor of $p_i$ for each incremental increase in $x$. 

\subsection{Denial of Service}

Since our protocol strictly interleaves PoW and PoS blocks, the network stalls if miners or stakers do not produce a block for some reason. This is a concern, but it is not significantly worse than a pure PoW network. 

\subsection{Selfish Mining}

The problem of \textit{selfish mining} in pure PoW networks is that miners who adopt this strategy hold onto blocks, mining from an advantageous position \cite{selfish_mining}. This is done until such a point that the network catches up to them, at which point, they publish their chain. This strategy wastes the efforts (work) of honest miners, thus netting the attacker a block production rate disproportionate to their actual mining power. Our protocol does not suffer from this problem specifically, since mining and staking blocks must be interleaved.

There is, however, a similar problem that our protocol is vulnerable to. A staker can collude with a miner by sending the miner a PoS block before the waiting time has elapsed. This miner now gets a head start on mining a PoW block on top of it. This costs the staker nothing (since she can still release her block to the wider network when it becomes eligible). It does give the miner the chance to win more of the blocks than their share of the hash power. Further analysis is required to define the bounds around the advantage a miner can achieve by participating in such a scheme.

%% file: sections/5results.tex
\section{Simulation Results}
\label{results}

We present empirical findings from both simulation and public test network (Amity) data to verify the behaviour of the algorithm. In the following presentation, we verify key system properties by corroborating simulation results with  s

e pick a property of the protocol that we're interested in verify and present simulation results to depict the behaviour. When applicable, we present results from the test network.

\subsection{Setup}

The simulation is set up as a game between two actors: the honest network, and
the attacker (in some setups). The game starts at a point in time given some initial state,
$I=(td_w,td_s,d_w,d_s)$. Where $td_w, td_s$ refers to the total staking or
working difficulty at block $n-1$, and $d_w, d_s$ refers to the last difficulty
or the PoW and PoS blocks.

For the purposes of some experiments, we generate the block generation scheme
for the honest party and observe results. For others, we simulate both parties. This
is done by allotting both parties with a time interval, and running a
random walk following the rules of the protocol. In scenarios where we want to
observe attacker success rate (for example, in the case of the double spend), we
repeatedly run a random walk to observe success rates at various power ratios.

\subsection{Block Generation Fairness}

We start with examining block generation fairness, by examining the base case of two players, an honest player with static hash rate and voting power, and an attacker with variable quantities. For this simulation we always assume the network is in a \textit{steady state}. We define a steady state as the following, given $H$ is the network total hashing power, and $V$ is the network total voting power. Then steady state is when these two variables are constant, and $d_s = V \cdot T$, $d_w = H \cdot T$.

We examine the block production with a time allotment of 30 days. Figure \ref{tab:table_block_fairness} summarizes the cases analyzed and simulation results. 

\begin{figure}[h]
    \centering
    \begin{tabular}{| c || c |}
        \hline
        \multicolumn{2}{| c |}{\textbf{Equal Power Scenario}} \\
        \hline
        Total Blocks Generated &  252575 \\
        Block Generation Ratio (Total) & 0.5000 \\
        Block Generation Ratio (PoS) & 0.5002 \\
        Block Generation Ratio (PoW) & 0.4997 \\
        \hline
        \multicolumn{2}{| c |}{\textbf{2x Hash Rate, No Voting Power Scenario}} \\
        \hline
        Total Blocks Generated & 252619 \\
        Block Generation Ratio (Total) & 0.3331\\
        Block Generation Ratio (PoS) & 0 \\
        Block Generation Ratio (PoW) & 0.6662 \\
        \hline
        \multicolumn{2}{| c |}{\textbf{2x Voting Power, No Voting Power Scenario}} \\
        \hline
        Total Blocks Generated & 253135 \\
        Block Generation Ratio (Total) & 0.3327 \\
        Block Generation Ratio (PoS) & 0.6654 \\
        Block Generation Ratio (PoW) & 0 \\
        \hline
    \end{tabular}
    \caption{Attacker block production under various scenarios. This data was
    generated at steady state conditions, with a time allotment of 30 days.}
    \label{tab:table_block_fairness}
\end{figure}

First, we examine the case where both players have equal hashing and voting power. In this case the expected rewards for both players should be identical. We see this is true in the \textit{equal power scenario} as block production for both sides are identical. In scenarios two and three, we study the case where the attacker has majority dominance in one of the two resources. The expectation is that one resource can at most only obtain 50\% of the rewards due to the interleaving structure of the blockchain. The percentage of blocks produced (at steady state) should on average be:

\begin{equation}
    P(win) = \frac{1}{2}v_i/\sum_i{v_i} + \frac{1}{2}h_i/\sum_i{h_i}.
\end{equation}

In both cases, the attacker produced blocks as predicted by the above equation.

\subsection{Block Time Distribution}

\begin{figure}[h]
    \centering
    \includegraphics[width=0.55\linewidth]{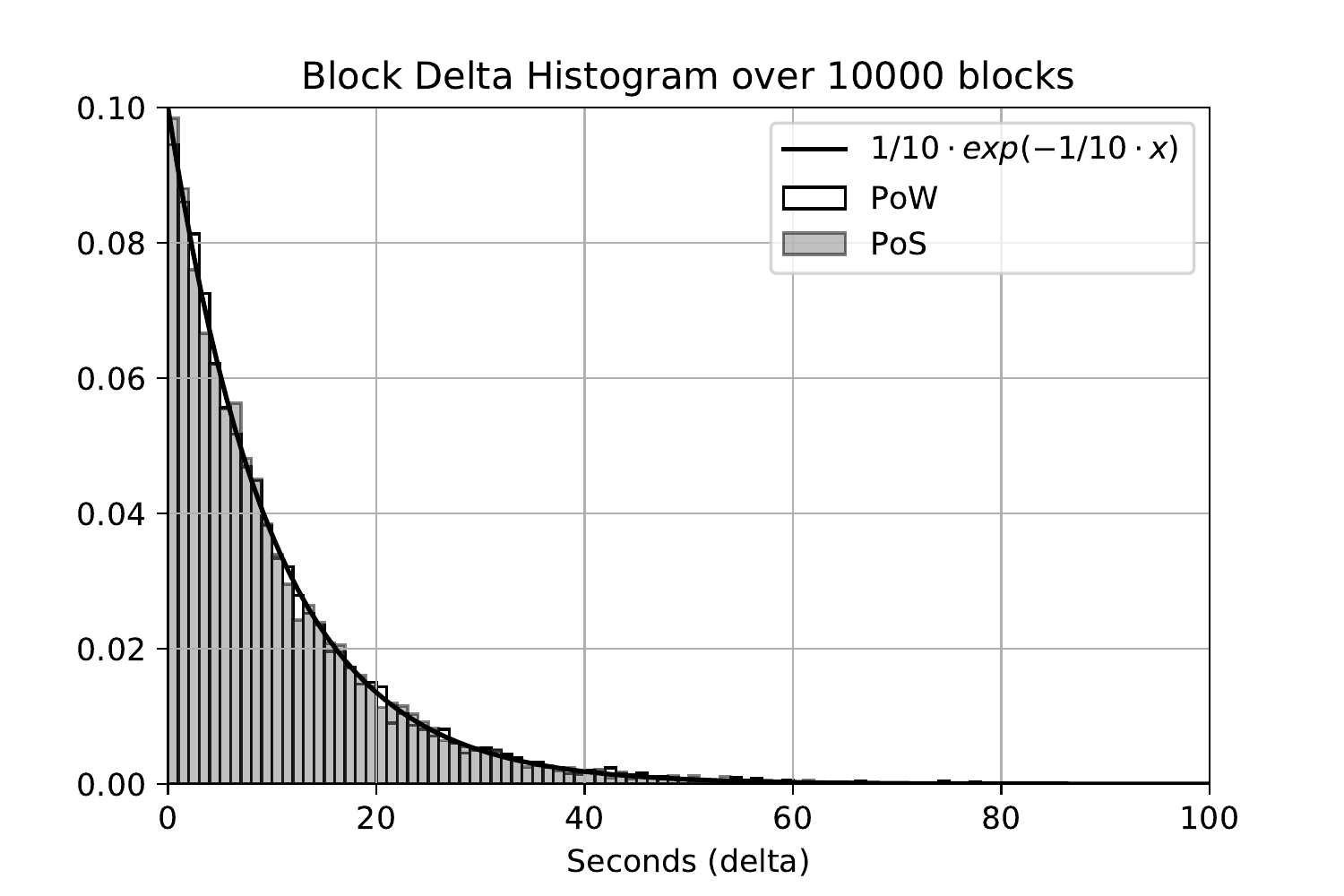}
    \caption{Histogram for PoW \& PoS block generation (100 bins with 1 second spacing), in \textit{steady-state} setting.}
    \label{fig:pos_pow_delta_histogram}
\end{figure}

Then, we look at the block time distribution based on our simulation. As described earlier, the protocol
difficulty adjustment function targets a mean block time of $T$ seconds, which is set in the simulations
and on Amity to be 10 seconds. Figure \ref{fig:pos_pow_delta_histogram} demonstrates that at steady state, the two generation algorithms exhibit extremely similar exponential PDFs, with $\lambda = \frac{1}{10}$ (rate) (as conjectured in \ref{difficulty_adjustment}). 

\begin{figure}[h]
    \centering
    \subfloat[PoW]{{\includegraphics[width=0.45\linewidth]{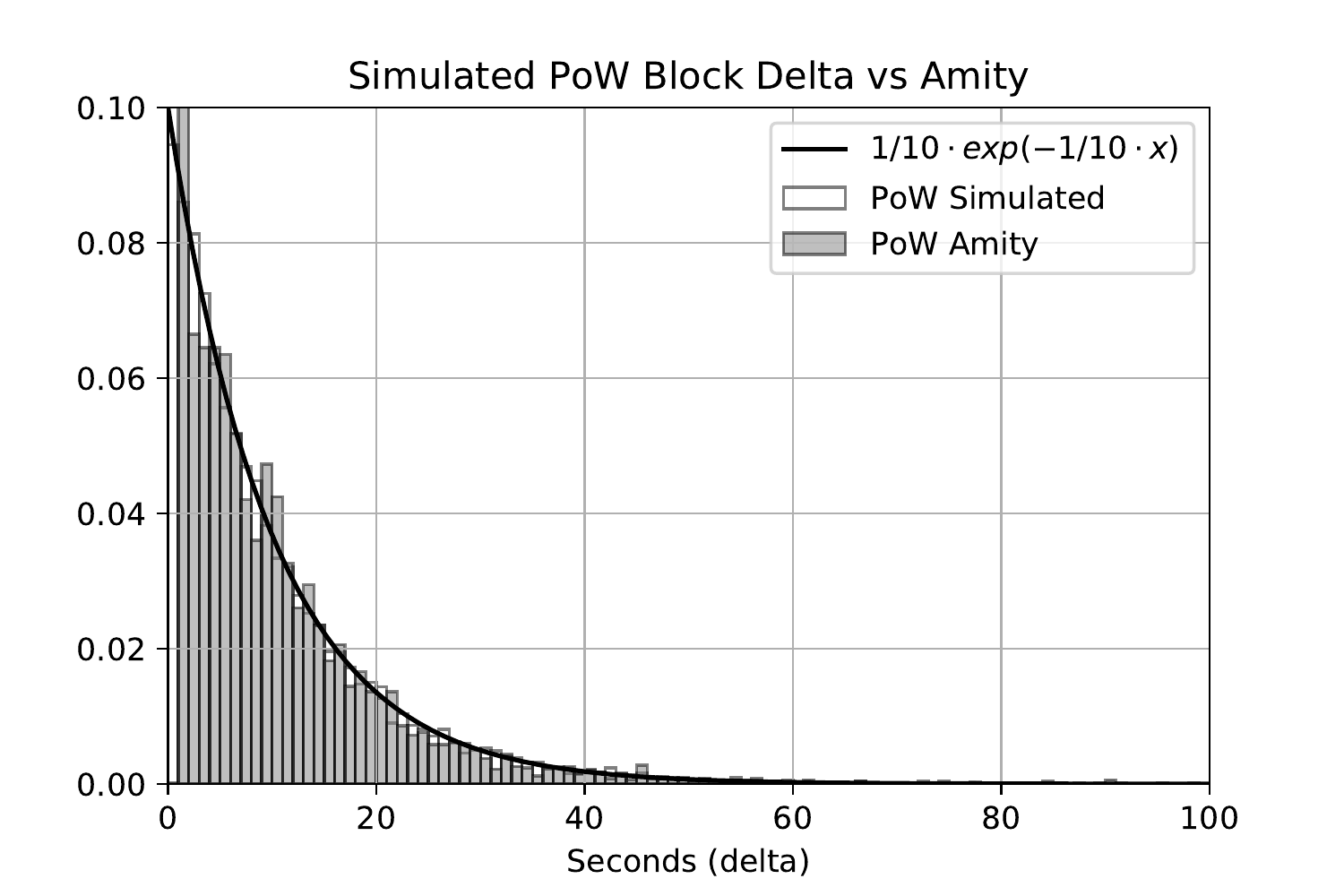} }}
    \qquad
    \subfloat[PoS]{{\includegraphics[width=0.45\linewidth]{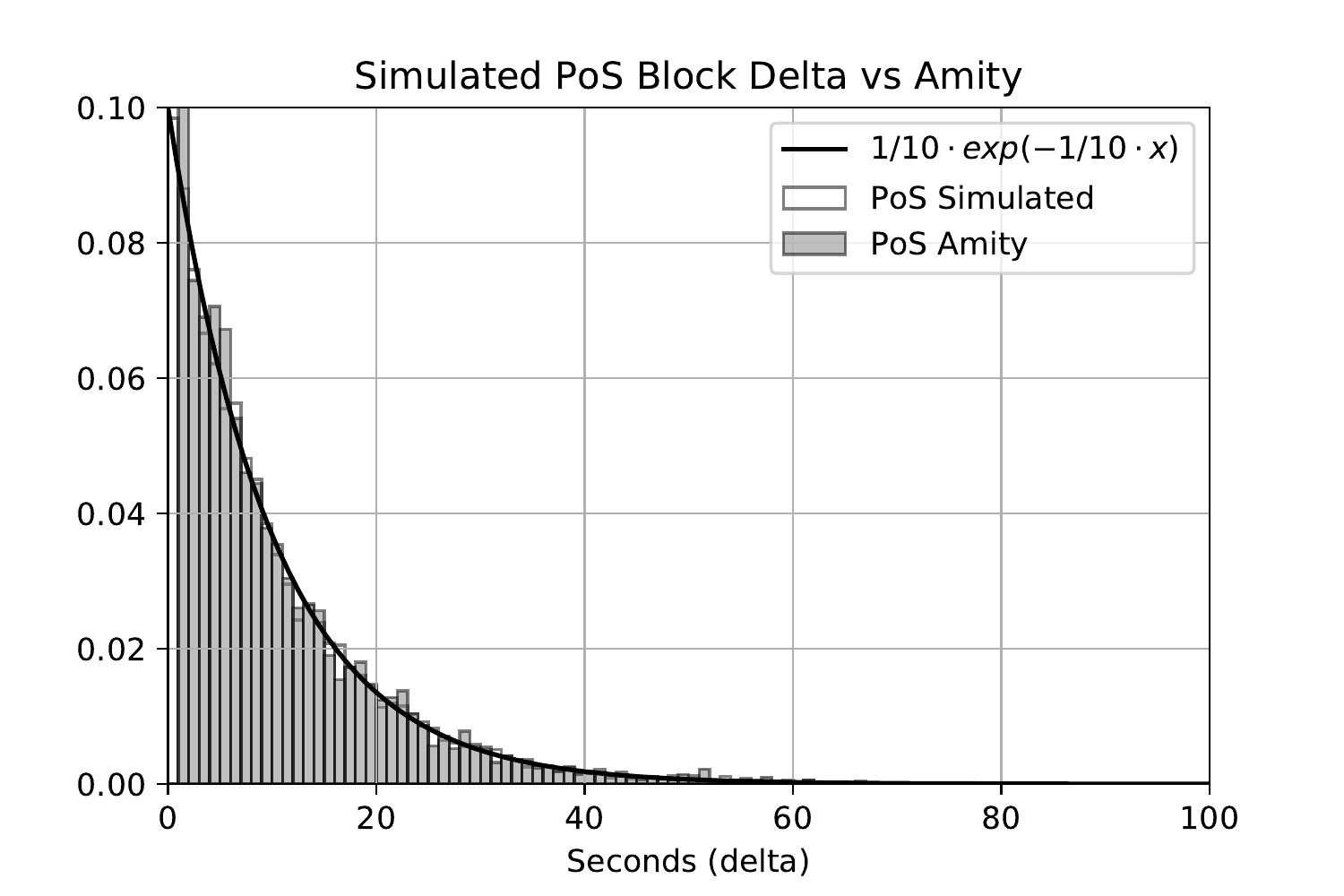} }}
    \caption{Density histogram of simulated block deltas, vs. data from Amity (10000 consecutive blocks).}
    \label{fig:delta_histogram}
\end{figure}

Figure \ref{fig:delta_histogram} demonstrates that simulation results are a fair representation of reality. We attribute the differences in results to the effects of network delay and propagation, and the fact that the sampling range included portions in which $V$ and $H$ fluctuated.

Also plotted on each graph is the theoretical distribution, the PDF of the exponential distribution with
the corresponding rate. It is obvious then that the expected mean and variance should be around 10 seconds.
The results listed in figure \ref{fig:block_time_mean_std} support this fact. We do note that there is higher variance on the live network, again this can be attributed to the increase in noise due to network propagation delays, and fluctuating resources.

\begin{figure}[h]
    \centering
    \begin{tabular}{| c || c | c |}
        \hline
        \multicolumn{3}{| c |}{\textbf{$E$ and $\sigma$ of Simulations and Amity}} \\
        \hline
        & Mean & Standard Deviation \\
        \hline
        Simulation PoW & 10.1965 & 10.6469 \\
        Simulation PoS & 10.1174 & 10.3223 \\
        Amity PoW & 9.6748 & 11.5929 \\
        Amity PoS & 9.8408 & 10.2822 \\
        \hline
    \end{tabular}
    \caption{Block production rate statistics in simulation and Amity test network settings.}
    \label{fig:block_time_mean_std}
\end{figure}

\subsection{Difficulty Adjustment}

\begin{figure}[h]
    \centering
    \includegraphics[width=0.55\linewidth]{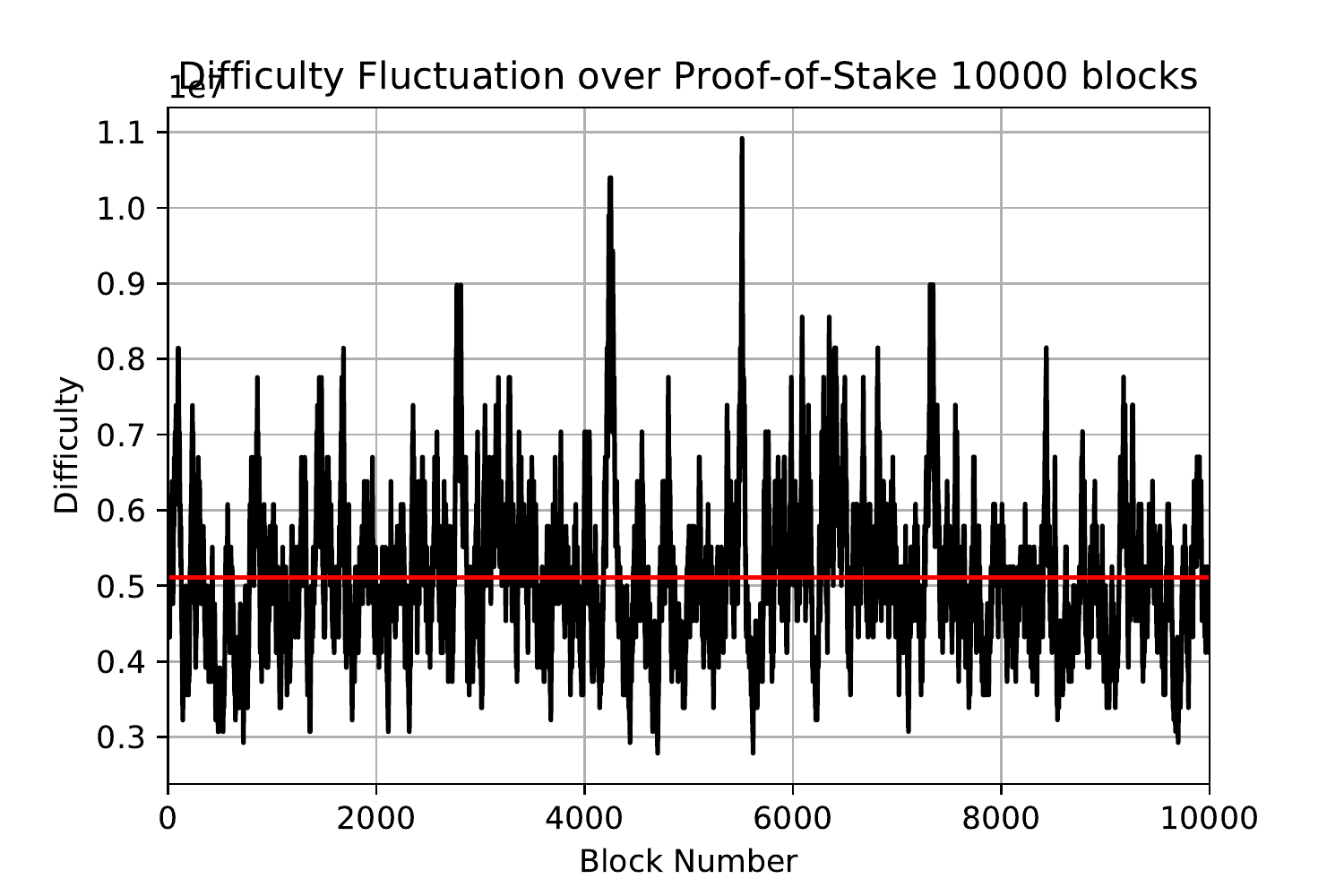}
    \caption{PoS difficulty over 10000 blocks. The simulation results represent
    a steady state (without perturbation of hashing or voting power). The parameters used
    were voting power = 500000, with an initial difficulty of $d_s=5000000$. Therefore we
    expect the difficulty (on average) to trend towards $d_s$, the resulting difficulty on
    average = 5109031.}
    \label{fig:pos_block_difficulty}
\end{figure}

\begin{figure}[ht]
    \centering
    \includegraphics[width=0.55\linewidth]{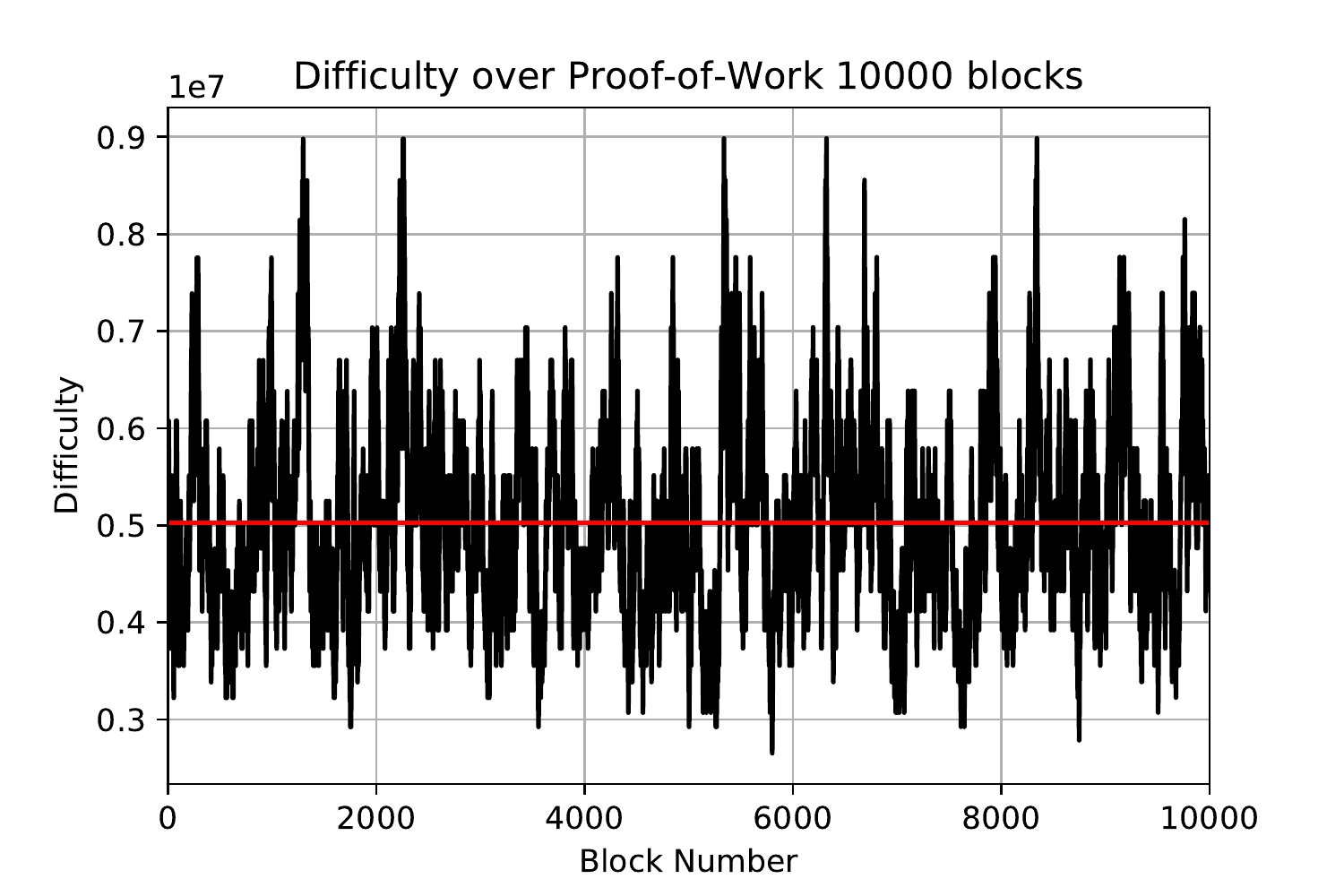}
    \caption{PoW difficulty over 10000 blocks, at steady state. The parameters used
    were hashing power = 500000, with an initial difficulty of $d_w=5000000$. Resulting average
    difficulty = 5027743.}
    \label{fig:pow_block_difficulty}
\end{figure}

Difficulty adjustment is tuned in the protocol to be more aggressive than the previous implementation,
that largely follows a retargeted version of the difficulty adjustment algorithm
introduced in \cite{wood2014ethereum}. Additionally, it is tuned specifically to target a
10 second block time, whereas the previous algorithm targeted a specific range. As depicted
in figures \ref{fig:pow_block_difficulty} and \ref{fig:pos_block_difficulty}, the 
adjustment algorithm can properly target the desired block time, and reaches
the expected difficulty of the network.

%% file: sections/6future-works-conclusion.tex
\section{Conclusions}

We provided an update to the original Unity protocol presented in \cite{wu2019unifying},
depicting the flaws in the protocol in \S\ref{double_spend_attack} and the motivation
that initially led to the creation of the Unity Interleave. We presented theoretical
and empirical evidence that the updated protocol produces a stable network, and is
resistant to a variety of attacks, benefiting from its hybrid nature.

There is more work to be done related to the protocol. Our exploration of attack vectors
while useful is still somewhat preliminary. As indicated in the original Unity paper, future work concerns deeper analysis of system stability under complex miner and staker dynamics and derivation of bounds for the protocol's security, availability and consistency. 